\input{aipcheck}

\documentclass[
    ,final            
  ]
  {aipproc}

\layoutstyle{6x9}

\begin{document}

\title{Searching for a Gravitational Heating Signature in Nearby
  Luminous Ellipticals}


\author{Tomer Tal}{
  address={Yale University}
}

\author{Pieter G. van Dokkum}{
  address={Yale University}
}

\author{Jeffrey D. P. Kenney}{
  address={Yale University}
}

\begin{abstract}
  We present a new deep optical study of a luminosity limited sample of
  nearby elliptical galaxies, attempting to observe the effects of
  gravitational interactions on the ISM of these objects.
  This study is motivated by recent observations of M86, a nearby
  elliptical galaxy that shows possible evidence for gas
  heating through a recent gravitational interaction.
  The complete sample includes luminous ellipticals in clusters,
  groups and the field.
  For each of the galaxies we objectively derive a tidal parameter which
  measures the deviation of the stellar body from a smooth, relaxed
  model and find that 73\% of them show tidal disturbance signatures in
  their stellar bodies.
  This is the first time that such an analysis is done on a
  statistically complete sample and it confirms that elliptical
  galaxies continue to grow and evolve through gravitational
  interactions even in the local Universe.
  Our study of ellipticals in a wide range of interaction stages,
  along with available ISM data will attempt to shed light on this
  possibly alternative mechanism for maintaining the observed ISM
  temperatures of elliptical galaxies.
\end{abstract}

\maketitle


\section{Introduction}
  Despite the relative simplicity of their stellar populations,
  elliptical galaxies vary significantly in their x-ray properties.
  The x-ray emission from giant ellipticals is dominated by radiation
  from a hot gaseous halo that engulfs the stars.
  Although gas is abundant in most of these systems, further star
  formation is suppressed by some mechanism which keeps the gas
  hot.
  
  Recently it has been suggested that active galactic nuclei
  are responsible for supplying energy to the hot x-ray halos
  \citep[e.g.][]{croton_many_2006} and for ionizing cold gas in their
  vicinity, thus creating the H$\alpha$ filaments that are often
  observed in elliptical systems
  \citep[e.g][]{mould_jet-induced_2000}.
  Alternatively, it was suggested that the source of these emission
  line filaments is cold gas accretion into the ISM of the
  galaxy by strong gravitational interactions and merger events
  \citep{sparks_plasmas_2004}.
  In this scenario, gas is stripped from
  the colliding galaxy and gets heated up, perhaps through heat
  conduction, by the x-ray halo of the elliptical.
  Recent analytical studies have shown that minor mergers and
  low-mass clumpy accretion may be sufficient to keep the halos of
  ellipticals hot \citep{dekel_gravitational_2008}.
  The interaction in this case transforms the initial potential
  energy of the accreted gas into thermal energy as it dissipates into
  the hot halo of the elliptical.
  The heating mechanism is therefore dynamical, consisting of weak
  shocks and/or ram-pressure drag to inject energy into both cold and
  hot interacting media.
  According to the gravitational heating model the interaction
  deposits energy into the ISM and heats it up,
  allowing for non AGN ellipticals to keep their x-ray
  halos hot.

\section{Evidence for Gravitational Heating}
\begin{figure}
    \includegraphics[width=0.57\textwidth]{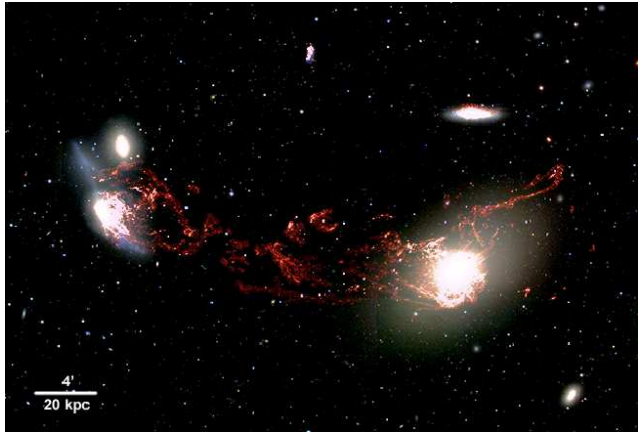}
    \caption{This continuum subtracted H$\alpha$+[NII] image of M86
    and NGC 4438 may provide the first observational evidence of gas
    heating through gravitational interactions.  The spiral is missing
    95\% of its cold gas.}
      \hfill
    \label{fig:m86}
  \end{figure}

  Suggestive evidence for gravitational heating comes from new
  observations of the Virgo galaxy M86 (figure \ref{fig:m86})
  \citep{kenney_spectacular_2008}, where we have discovered a
  remarkable set of large scale H$\alpha$+[NII] filaments connecting
  the giant elliptical to NGC 4438, a spiral galaxy $\sim$23'
  ($\sim$120 kpc) away.
  Spectroscopy of the filaments show a fairly smooth velocity
  gradient between the galaxies, strongly suggesting
  a recent gravitational interaction between them.
  The data suggest that as NGC 4438 passed by M86 at high velocity,
  its cold gas was probably stripped by the ISM of the elliptical,
  causing the spiral to become HI-deficient.
  Despite being x-ray bright, M86 shows no clear signs for an active
  nucleus.
  Moreover, the total kinetic energy of the cold gaseous component of
  the spiral at closest approach was comparable to the current total
  thermal energy stored in the hot halo of the elliptical.

\section{OBEY: Observations of Bright Ellipticals at Yale}
  We present a deep broadband optical imaging study of a complete
  sample of luminous elliptical galaxies ($M_B<-20$) at distances 15
  Mpc - 50 Mpc, selected from the Tully catalog of nearby galaxies.
  The images are flat to $\sim$0.35\% across the 20' field and reach
  a V band depth of 27.7 mag arcsec$^{-2}$.
  We derive an objective tidal interaction parameter for all galaxies
  and find that 73\% of them show tidal disturbance signatures in
  their stellar bodies, in agreement with the findings of van Dokkum
  (2005) \cite{dokkum_recent_2005} who studied a sample of red
  galaxies at $z\sim0.1$.
  This is the first time that such an analysis is done on a
  statistically complete sample and it confirms that tidal features in
  ellipticals are common even in the local Universe
  \cite{tal_frequency_2009}.

  By comparing the tidal interaction parameter of the galaxies to
  their broad band optical colors we find that interacting systems are
  slightly bluer than non interacting ones, implying that these
  mergers are accompanied by little or no star formation.
  In fact, it appears that even the most interacting galaxies in our
  sample are red.
  This likely suggests that either very small amounts of cold gas
  are driven into the elliptical by the interaction or that star
  formation is suppressed as the accreted gas is heated and ionized by
  the collision.

  \begin{figure}
    \includegraphics[width=0.49\textwidth]{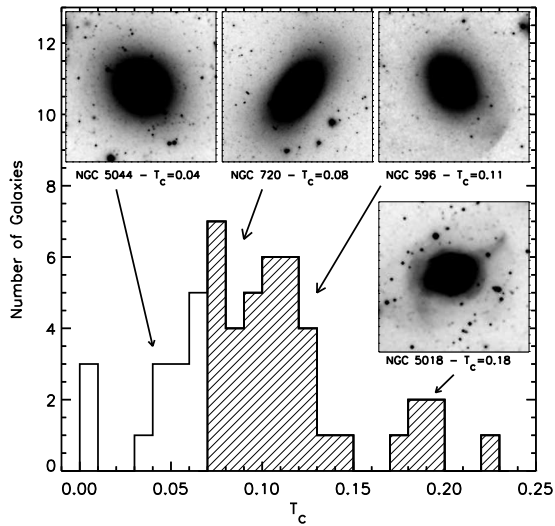}
    \caption{The distribution of derived tidal parameter values. The
      shaded area represents galaxies with a T$_c$ value greater than
      the detection threshold. The four subset images are typical
      examples of various interaction levels.}
      \hfill
    \label{fig:tp_distr}
  \end{figure}

\section{Tidal Features and the AGN Duty Cycle}
  It has been suggested by several authors that gravitational
  interactions play an important role both in evolving elliptical
  galaxies \citep[e.g.][]{kormendy_structure_2009} and in triggering
  galactic nuclear activity in them. 
  This is shown by a correlation between the brightness of
  elliptical galaxies in radio continuum observations and
  morphological disturbances
  \citep[e.g.][]{heckman_galaxy_1986,smith_multicolor_1989}. 

  We find no relation between the tidal parameter and radio continuum
  flux of our survey galaxies.
  Since the lifetime of the radio mode is only roughly $10^8$ years
  \citep[e.g.][]{croton_many_2006,shabala_duty_2008}, galaxies with a
  low tidal parameter are all expected to be quiet at radio
  wavelengths.
  The existence of radio-loud AGN in undisturbed ellipticals is very
  interesting and may imply that gravitational interactions are not
  the only and possibly not the most important AGN triggering
  mechanism.

\section{Gravitational Interactions and ISM Heating}
  Similarly to other ISM heating mechanisms gravitational interactions
  can potentially inject significant and sufficient amounts of energy
  into the hot halos of massive ellipticals.
  Simplified calculations, such as were demonstrated by Dekel and
  Birnboim (2008) show that an accretion rate of only $10^{-3}-10\ 
  M_{\odot}\ yr^{-1}$ would be needed to support a typical nearby
  halo.
  We have shown that the frequency of such collisions is high even in
  the local Universe and that more than two-thirds of all nearby
  massive ellipticals show disturbance signatures in their stellar
  bodies.
  However, not unlike other proposed ISM heating mechanisms (e.g.,
  AGN, conduction, sloshing) the gravitational heating model lacks
  details regarding the specific energy transfer processes at work.
  In summary, although not a strong evidence for the importance of
  gravitational heating by itself, the implied interaction rate from
  our study suggests that mergers and accretion events are frequent
  enough to support such a model.

  \begin{figure}
    \includegraphics[width=0.44\textwidth]{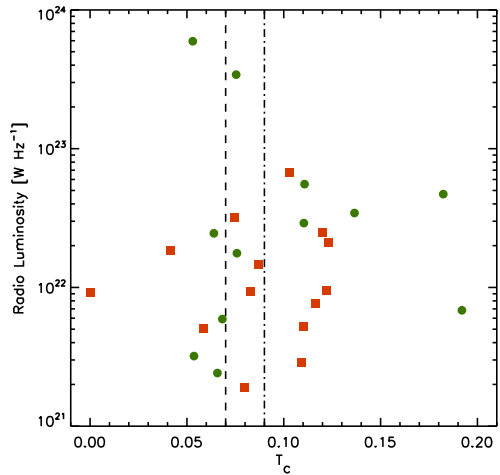}
    \caption{The relation between radio continuum flux and tidal
      parameter for 26 of the sample galaxies.
      The existence of radio-load, non-interacting galaxies suggests
      that gravitational interactions are not the only and possibly
      not the most important AGN triggering mechanism.}
    \hfill
    \label{fig:radiotp}
  \end{figure}





\bibliographystyle{aipproc}   

\bibliography{tal_t}

\begin{thebibliography}{11}
\expandafter\ifx\csname natexlab\endcsname\relax\def\natexlab#1{#1}\fi
\providecommand{\enquote}[1]{``#1''}
\expandafter\ifx\csname url\endcsname\relax
  \def\url#1{\texttt{#1}}\fi
\expandafter\ifx\csname urlprefix\endcsname\relax\def\urlprefix{URL }\fi
\providecommand{\eprint}[2][]{\url{#2}}

\bibitem[Croton et~al.(2006)]{croton_many_2006}
D.~J. Croton, V.~Springel, S.~D.~M. White, G.~D. Lucia, C.~S. Frenk, L.~Gao,
  A.~Jenkins, G.~Kauffmann, J.~F. Navarro, and N.~Yoshida, \emph{Monthly
  Notices of the Royal Astronomical Society} \textbf{365}, 11--28 (2006).

\bibitem[Mould et~al.(2000)]{mould_jet-induced_2000}
J.~R. Mould, A.~Ridgewell, J.~S. Gallagher, M.~S. Bessell, S.~Keller,
  D.~Calzetti, J.~T. Clarke, J.~T. Trauger, C.~Grillmair, G.~E. Ballester,
  C.~J. Burrows, J.~Krist, D.~Crisp, R.~Evans, R.~Griffiths, J.~J. Hester,
  J.~G. Hoessel, J.~A. Holtzman, P.~A. Scowen, K.~R. Stapelfeldt, R.~Sahai,
  A.~Watson, and V.~Meadows, \emph{Astrophysical Journal} \textbf{536},
  266--276 (2000).

\bibitem[Sparks(2004)]{sparks_plasmas_2004}
W.~B. Sparks, \enquote{Plasmas in Galaxies: Ionized Gas in Elliptical
  Galaxies,} 2004, vol. 703, pp. 291--299.

\bibitem[Dekel and Birnboim(2008)]{dekel_gravitational_2008}
A.~Dekel, and Y.~Birnboim, \emph{Monthly Notices of the Royal Astronomical
  Society} \textbf{383}, 119--138 (2008).

\bibitem[Kenney et~al.(2008)]{kenney_spectacular_2008}
J.~D.~P. Kenney, T.~Tal, H.~H. Crowl, J.~Feldmeier, and G.~H. Jacoby,
  \emph{Astrophysical Journal} \textbf{687}, L69--L74 (2008).

\bibitem[van Dokkum(2005)]{dokkum_recent_2005}
P.~G. van Dokkum, \emph{Astronomical Journal} \textbf{130}, 2647--2665 (2005).

\bibitem[Tal et~al.(2009)]{tal_frequency_2009}
T.~Tal, P.~G. van Dokkum, J.~Nelan, and R.~Bezanson, \emph{Astronomical
  Journal} \textbf{138}, 1417--1427 (2009).

\bibitem[Kormendy et~al.(2009)]{kormendy_structure_2009}
J.~Kormendy, D.~B. Fisher, M.~E. Cornell, and R.~Bender, \emph{Astrophysical
  Journal Supplement Series} \textbf{182}, 216--309 (2009).

\bibitem[Heckman et~al.(1986)]{heckman_galaxy_1986}
T.~M. Heckman, E.~P. Smith, S.~A. Baum, W.~J.~M. van Breugel, G.~K. Miley,
  G.~D. Illingworth, G.~D. Bothun, and B.~Balick, \emph{Astrophysical Journal}
  \textbf{311}, 526--547 (1986).

\bibitem[Smith and Heckman(1989)]{smith_multicolor_1989}
E.~P. Smith, and T.~M. Heckman, \emph{Astrophysical Journal} \textbf{341},
  658--678 (1989).

\bibitem[Shabala et~al.(2008)]{shabala_duty_2008}
S.~S. Shabala, S.~Ash, P.~Alexander, and J.~M. Riley, \emph{Monthly Notices of
  the Royal Astronomical Society} \textbf{388}, 625--637 (2008).

\end{thebibliography}

\end{document}